\documentclass[aps,prl,twocolumn,showpacs,10pt,preprintnumbers,nofootinbib,reprint,groupedaddress,superscriptaddress]{revtex4-1}

\usepackage{amsmath}
\usepackage{amsfonts}
\usepackage{amssymb}
\usepackage{graphicx, rotating}
\usepackage{epstopdf}
\usepackage{epsfig}
\usepackage{latexsym}
\usepackage{mathtools}
\usepackage{graphicx}
\usepackage{color}
\usepackage[dvipsnames]{xcolor}
\usepackage{amsmath,amssymb}
\usepackage{multirow}
\usepackage{array,makecell}
\usepackage[utf8]{inputenc}
\usepackage{arydshln}
\usepackage{ mathrsfs }
\usepackage{tabularx}
\usepackage{ bbold }

\usepackage{tikz-feynman}
\tikzfeynmanset{warn luatex=false}

\usepackage{slashed}
\usepackage{hyperref}
\hypersetup{colorlinks, citecolor=bluscuro, linkcolor=bluscuro, urlcolor=bluscuro}
\definecolor{rossos}{cmyk}{0,1,1,0.55}
\definecolor{bluscuro}{rgb}{0.15, 0.2, .85}
\definecolor{bluchiaro}{cmyk}{1,.3,0.,0.1}
\definecolor{verdescuro}{rgb}{0.3,0.8,0.3}

\newcommand{\be}{\begin{equation}}
\newcommand{\ee}{\end{equation}}          
\newcommand{\bea}{\begin{eqnarray}}
\newcommand{\eea}{\end{eqnarray}}
\newcommand{\bc}{\begin{center}}
	\newcommand{\ec}{\end{center}}

\newcommand{\op}{\,\mathcal{O}}

\def\cale{{\mathcal{E}}}

\def\MG{\texttt{MadGraph}}
\def\PT{\textsc{Pythia}8}

\newcommand{\ab}[1]{\langle #1 \rangle}

\newcommand{\la}[1]{\langle #1 |}

\newcommand{\ra}[1]{| #1 \rangle }

\begin{document}

\preprint{CERN-TH-2024-113}

\title{One-point correlators of conserved and non-conserved charges in QCD}

\author{Marc Riembau}
\affiliation{Theoretical Physics Department, CERN, 1211 Geneva 23, Switzerland}

\author{Minho Son}
\affiliation{Department of Physics, Korea Advanced Institute of Science and Technology,
	291 Daehak-ro, Yuseong-gu, Daejeon 34141, Republic of Korea}

\begin{abstract}
\noindent 
{
One-point correlators of conserved charges are argued to be perturbatively IR safe in QCD, 
which includes not only the density of energy, but also those of electric charge, isospin and baryon number.
Theoretical and phenomenological aspects of the density matrix of one-point correlators are discussed in the context of the states produced by a chiral current, as in the decay of a polarized electroweak boson. 
Densities of some non-conserved charges such as energy with arbitrary non-negative powers, despite their incalculability, are shown to obey an infinite set of consistency constraints. 
QCD is observed to live near a kink in the allowed parameter space of one-point correlators.
}

\end{abstract}

\maketitle

\medskip

\section{Introduction}

The S-matrix in QCD is a theoretically well-defined observable thanks to the mass gap. However, due to the large multiparticle production, the measurement of the production rate of each individual state is unfeasible.
It is convenient therefore to \textit{coarse grain} the Hilbert space and measure instead the cross section to produce an ensemble of states, namely jets of particles \cite{PhysRevLett.39.1436}, integrating over the individual constituents of those.

Another avenue for extracting useful information of the underlying physics is to measure the correlation among energy fluxes.
These are a set of event shape observables proposed long ago in the context of $e^+e^-$ annihilations \cite{PhysRevD.17.2298,PhysRevLett.41.1585,PhysRevD.19.2018,ORE198093},
later connected with correlators of the stress-energy tensor \cite{Sveshnikov:1995vi,Cherzor:1997ak,Korchemsky:1997sy,Belitsky:2001ij,Craft:2022kdo} and observed to obey an operator product expansion \cite{Hofman:2008ar}. 
Recent analytic progress of energy correlators in QCD and $\mathcal{N}=4$\,SYM appeared in 
\cite{Belitsky:2013xxa,Belitsky:2013bja,Belitsky:2013ofa,Dixon:2018qgp,Henn:2019gkr,Korchemsky:2019nzm,Kologlu:2019mfz}.
While event shape observables do not depend on a jet algorithm, looking into specific patterns of energy fluxes within an energetic jet at the LHC has proven powerful for understanding QCD \cite{Chen:2020vvp,Komiske:2022enw,Chen:2023zlx,CMS:2024mlf}. 

In this \textit{Letter}, we initiate a phenomenological exploration of correlators of conserved charges and discuss a theoretical connection with a class of non-conserved ones. 
Charge correlators have been considered in $\mathcal{N}=4$\,SYM and QCD at leading order \cite{Hofman:2008ar,Belitsky:2013xxa,Belitsky:2013bja,Belitsky:2013ofa,Chicherin:2020azt}. At the phenomenological level, a related quantity is given by the jet charge as defined in \cite{FIELD19781}, given by $Q_\kappa = \sum_{i\in \text{Jet}}Q_i p_{T,i}^\kappa$. The energy weight makes the assigned jet charge stable against the radiation of soft particles. This definition of the jet charge has been extensively used by experimental collaborations \cite{Erickson:1979wa,EuropeanMuon:1984xji,DELPHI:1991mqi,ALEPH:1991fba,OPAL:1992jsm,ATLAS:2015rlw,CMS:2017yer,CMS:2020plq} and explored theoretically
\cite{Krohn:2012fg,Waalewijn:2012sv,Elder:2017bkd,Chen:2019gqo,Li:2019dre,Kang:2020fka,Kang:2021ryr,Jaarsma:2022kdd,Lee:2022kdn,Kang:2023ptt}.
In this \textit{Letter}, we explore charge correlators as a way to study the charge density of a chiral current. While $Q_\kappa$ is defined event-by-event and the measurement aims to extract the production rate of different values of $Q_\kappa$, correlation functions work in the opposite way and are defined as an average over an ensemble of events. We argue that it is this fact what makes them infrared-safe (IR-safe) perturbatively. In QCD, this includes the energy density, but also the one-point correlators of the other global charges: electric charge, isospin and quark flavors, and baryon number. 
We compute such densities perturbatively and explore non-perturbative corrections due to the hadronization and weak interactions, and observe how the different densities carry an incredible amount of information about the dynamics despite the simplicity of the observable. 
We comment on when higher point correlators are also IR-safe. 

We show that Lorentz invariance, unitarity and the positivity of the energy lead to stringent constraints on the analytic structure and behavior of the one-point correlator. While it includes the bounds of \cite{Hofman:2008ar}, our approach provides a detailed view on it, and  trivially extends to non-conserved fluxes like $\cale^k$ for $k\neq 1$. We further present the complete and optimal set of constraints that the densities among different $k$ must obey. 
An interesting application is to constrain experimentally challenging observables with low $k$, like the particle number density, from experimentally clean observables with a large $k\gtrsim 1$.
We discovered that, when extrapolating to low $k$, the numerical simulation of the data sits close to a kink in the allowed parameter space. We interpret this as an indication that the distribution in $k$ is close to a power law.

\section{Densities of conserved charges}

Measuring the asymptotic charge of a conserved current means to place a detector $\mathcal{D}_{n}$ at spatial infinity in some direction $\vec{n}$, which can be thought of as an operator \cite{Sveshnikov:1995vi,Cherzor:1997ak,Korchemsky:1997sy} acting on physical multiparticle states $|\alpha\rangle$ as
\be
\mathcal{D}_{n} |\alpha\rangle \,=\, \sum_i \omega_i \delta^{(2)}(\Omega_{n}-\Omega_i) |\alpha\rangle\, ,
\ee
where $\Omega_i$ is the infinitesimal solid angle in the direction of $p_i$ and $\omega_i$ the quantum number of the particle $i$ measured by the detector, that is, the detector measures the quantum number $\omega_i$ of the particle $i$ of the multiparticle state $|\alpha\rangle$ only if the 3-momentum direction coincides with $\vec{n}$.
In the presence of an operator $\op$ exciting the QCD vacuum, the expected average of measurements of the detector, namely the one-point correlator, can be extracted from the three point correlator $\langle \op^\dagger \mathcal{D}_n \op\rangle$ as
\be
\ab{\mathcal{D}_n} = \frac{1}{\mathcal{N}}\int d^4 x e^{i p \cdot x}  \langle \op^\dagger(x) \mathcal{D}_n \op(0)\rangle\,~,
\ee
with the normalization $\mathcal{N}=\int d^4 x e^{i p \cdot x}  \langle \op^\dagger(x) \op(0)\rangle$. By inserting a complete set of states, it is clear that this measures the average charge in the direction $\vec{n}$ in the presence of a source $\op(x)$,
\be
\ab{\mathcal{D}_n} = \frac{1}{\mathcal{N}}\sum_{\alpha,i}\delta^{(d)}(p-p_\alpha) \, \omega_i \delta^{(2)}(\Omega_{n}-\Omega_i) 
|\la{\alpha} \op\ra{0}|^2\,~,
\ee
where $\ra{\alpha}$ runs over all multiparticle states in the theory, and $i$ runs over all particles in each state.

\subsection{IR safety of charge densities}

The fact that $\mathcal{D}_{n}$ measures a conserved quantity implies two important properties: \textit{i)} IR finiteness against soft and collinear radiation and \textit{ii)} connection between different phases of the theory. We explore both aspects in the following.

In gauge theories, exclusive rates to produce a single state have IR divergences signaling that the theory has nontrivial IR dynamics. Such divergences, via the KLN theorem \cite{Kinoshita:1962ur,Lee:1964is}, are canceled after summing over a set of degenerate states.
This suggests an introduction of IR safe observables as those that do not spoil the cancellation of IR divergences and therefore are dominated by the short distance dynamics \cite{PhysRevLett.39.1436,PhysRevD.46.192,Banfi:2010xy}. We now argue that one-point correlators of conserved charges belong to this category.

Since the conserved charge is preserved under collinear radiation, the detector $\mathcal{D}_{n}$ acts homogeneously on the set of particles collinear to the direction $\vec{n}$ measuring their total charge.
Moreover, the correlator is inclusive in the other collinear sectors, ensuring the cancellation of IR divergences due to the collinear radiation.

Soft radiation, instead, is annihilated by the detector of conserved charges.
Soft photons and gluons couple universally to the rest of the amplitude \cite{PhysRev.135.B1049}. This implies that the one-point correlator receives a contribution from an off-shell photon of a momentum $q^2$ that splits into a $q\bar{q}$ pair proportional to
\be
\int d\Phi_2 \frac{p_q^\mu p_{\bar{q}}^\nu+p_q^\nu p_{\bar{q}}^\mu-q^2/2\eta^{\mu\nu}}{q^2}\sum_{i=q\bar{q}} w_i \delta^{(2)}(\Omega_i-\Omega_n)\,,
\label{eq:softcontribution}
\ee
which vanishes since the integrand is symmetric under $p^\mu_q\leftrightarrow p^\mu_{\bar{q}}$ due to the vector-like nature of the gauge boson coupling and $w_q=-w_{\bar{q}}$ due to gauge bosons carrying no charge.
While at the event-by-event level the detector does receive contributions from soft particles, those average out in an ensemble of events.
The fact that the correlator computes the expected average over an inclusive set of states is what makes it insensitive to the soft radiation and to the appearence of non-global logarithms 
\cite{Dasgupta:2001sh,Banfi:2002hw}.

A second aspect that highlights the importance of considering detectors of conserved charges is the fact that, operatorially, those are defined as
\be
\mathcal{D}_n \,=\, \lim_{r\to\infty}\int_0^\infty dt\,r^2 n_i  J^i(t,r\vec{n})\, ,
\ee
where $J^\mu$ is a conserved current, given by $T^{0\mu}$ or $q_f\bar{f}\gamma^\mu f$ for a detector measuring the energy or the charge $q_f$ of a fermion. In QCD, the fact those are conserved currents implies that the detector can be expressed and measured in terms of hadrons, while at the same time can be expressed and computed in terms of quarks and gluons.
If we were to measure instead a non-conserved quantity, e.g. the energy squared, this would be associated to an operator which develops an anomalous dimension \cite{Hofman:2008ar},
and generically would be mapped to an unkown operator at parton level.

\subsection{Beyond one-point correlators}

The arguments for the correlator to be IR-safe perturbatively are not necessarily constrained to one-point correlators. 
As long as the detectors $\mathcal{D}_{n_i}$ consist of conserved charges, the correlator $\ab{\mathcal{D}_{n_1}\cdots \mathcal{D}_{n_m}}$ for well separated directions $\vec{n}_1,\dots , \vec{n}_m$ is collinear safe. 
 Each detector acts on a different collinear sector. By the same reasoning described above, they act homogeneously measuring the total charge on each sector and it is collinear safe.
 The fact that the detector measures a conserved charge is crucial, in contrast to a generic detector as explored in e.g. \cite{Lee:2023npz}.

IR-safety under soft emissions puts instead restrictions on the type of detectors, as they are forced to annihilate the soft sector, similar to Eq.~\ref{eq:softcontribution} but with a weight proportional to $w_q w_{\bar{q}}$ to signal that each detector measures a quark from the soft splitting.
This is annihilated whenever one of the detectors measures the energy, or when one of the detector measures a charge that is not produced or highly suppressed during the perturbative evolution, like $c$- and $b$-flavor. 

In this way, for instance, the electric charge-electric charge $\ab{\mathcal{Q}_{n_1}\mathcal{Q}_{n_2}}$ or isospin-baryon number $\ab{\mathcal{I}_{n_1}\mathcal{B}_{n_2}}$ correlators are sensitive to the soft radiation. 
Examples of IR-soft safe correlators are the energy-electric charge $\ab{\mathcal{E}_{n_1}\mathcal{Q}_{n_2}}$ or in the case of jets of heavy quarks, charm-isospin $\ab{c_{n_1}\mathcal{I}_{n_2}}$, or $b$-flavour - baryon number $\ab{b_{n_1}\mathcal{B}_{n_2}}$.
Once those are shown to be IR safe, any higher point correlator where all pairs are IR-safe will be IR-safe as well, e.g. $\ab{\mathcal{E}_{n_1}\mathcal{E}_{n_2}\mathcal{Q}_{n_3}}$.

This opens up a large, unexplored space of IR-safe correlators. An interesting system to compute and measure those is a top quark decay, since the chiral nature of the top and the richness of its decays leads to nontrivial correlations among the different quantum numbers. 
We leave this to future explorations.

\section{Charge densities
\\
of A Chiral Current}

In the following we focus on the simplest phenomenologically relevant operator that can excite the QCD vacuum, given by a chiral fermion current,
\be
J_{h}(x)\,=\, \varepsilon^{\mu}_h  \bar{q}\gamma_{\mu} P_{L} q   (x)\, ,
\ee
where $P_{L}$ gives the left and right chiral fermion current.
To simplify notation we consider a left handed current unless otherwise specified, and comment the difference from a right handed current.
The projection on the polarization vectors $\varepsilon^{\mu}_h$ allows us to write a density matrix, that will become relevant for collider quantities,
\begin{align}
	\begin{aligned}
		\langle \mathcal{D}_n\rangle_{hh^\prime} = \frac{1}{\mathcal{N}} \int d^4 x e^{i p \cdot x}  \langle J^\dagger_{h^\prime}(x) \mathcal{D}_n J_{h}(0)\rangle
	\end{aligned}\, ,
	\label{eq:densitymatrix}
\end{align}
with the normalization given by tracing over polarizations of the matrix element without a detector $\mathcal{N} = -\int d^4 x e^{i p \cdot x} \langle J^{\mu\dagger}(x) J_\mu(0)\rangle$, equal to the inclusive production of hadrons from $J^\mu$.
This density matrix represents the average charge 
measured in the direction $\vec{n}$ in the presence of the decay of a massive vector with a momentum $p^\mu$ and a helicity of $h,h^\prime=+,-,0$.

The three point $J\mathcal{D}J$ correlator,
\be
H^{\mu\nu}_\mathcal{D}\,=\,\int d^4 x e^{i p \cdot x} \langle J^{\mu\dagger}(x) \mathcal{D}_n J^\nu(0)\rangle\, ,
\label{eq:JDJcorrelator}
\ee
can be decomposed in terms of three independent tensor structures that depend on the injected momentum $p^\mu$ and the null detector's direction $n_\mu$ as
\be
\begin{split}
H^{\mu\nu}_\mathcal{D}&=\left( -\eta^{\mu\nu} +  p^\mu p^\nu/Q^2 \right) H_\mathcal{D}^0 
\\[5pt]
&+ \left( \frac{Q^2 n_\mu n_\nu +p_\mu p_\nu}{Q n\cdot p}-\frac{n_\mu p_\nu +p_\mu n_\nu}{Q} \right) H_\mathcal{D}^n
\\[5pt]
&-  \frac{i}{Q} \epsilon^{\mu\nu\alpha\beta}p_\alpha n_\beta H_\mathcal{D}^o\, ,
\label{eq:Hmunudecompositionfull}
\end{split}
\ee
where $H_\mathcal{D}^0$, $H_\mathcal{D}^n$, and $H_\mathcal{D}^o$ are generically functions of $p^2$ and $p\cdot n$.
The different components may be extracted by computing the contractions of $H^{\mu\nu}$ with $\eta_{\mu\nu}$, $n_\mu n_\nu$ and $\epsilon_{\mu\nu ab}p^a n^b$.
Notice that the scalars $H_\mathcal{D}(p^2,p\cdot n)$ are Lorentz-covariant and inherit the transformation properties of $\mathcal{D}$. 
We are interested in the dependence on the direction $\vec{n}$ of the detector.
For a generic $p^\mu$ and a generic direction $n^\mu$, $p\cdot n$ does depend on the polar and azimuthal angles $\theta$ and $\phi$.

In the rest frame, $p^\mu=(Q,\vec{0})$ and $n_\mu=(1,\vec{n})$, so that $p\cdot n=\sqrt{p^2}$ and the dependence on $\vec{n}$ is entirely on the tensor structure.
Contracting $H^{\mu\nu}$ with $\epsilon^\mu_h \epsilon^{*\nu}_{h^\prime}$, leads to $\delta_{hh^\prime}$, $(\vec{\epsilon}_h\cdot \vec{n})(\vec{\epsilon}^*_{h^\prime}\cdot \vec{n})$ and $-i\epsilon^{ijk}\epsilon_{h^\prime,i}^*n_j\epsilon_{h,k}$ from the first, second, and third row of Eq.~\ref{eq:Hmunudecompositionfull}, respectively.
The most general form that the density matrix of correlators can take is given by
\bea\nonumber
\langle \mathcal{D}_n\rangle_{hh^\prime} &=& \frac{\ab{\mathcal{D}}}{4\pi}\bigg[
\delta_{hh^\prime}+a_\mathcal{D} \left((\vec{\epsilon}_h\cdot \vec{n})(\vec{\epsilon}^*_{h^\prime}\cdot \vec{n})-\frac{\delta_{hh^\prime}}{3}\right)\bigg] \\
&& -\frac{i}{4\pi}\, b_\mathcal{D}\,\epsilon^{ijk}\epsilon_{h^\prime,i}^*n_j\epsilon_{h,k}\,,
\label{eq:chargedensityparametrization}
\eea
where $\ab{\mathcal{D}}$ is the total charge of the operator. The parameters $a_\mathcal{D}$ and $b_\mathcal{D}$ control the parity-even and parity-odd parts of the correlator, are determined by the dynamics of the theory and are independent of the helicity projected from the operator.
In terms of the Lorentz structures in Eq.~\ref{eq:Hmunudecompositionfull}, the overall normalization is given by $\mathcal{N} = 4\pi(H^0_{\mathbb{1}}+\frac13 H^n_{\mathbb{1}})\equiv 4\pi \mathscr{N}$ where $\mathbb{1}$ denotes the insertion of the identity in place of a detector, that is, it computes the inclusive production rate. The coefficients in Eq.~\ref{eq:chargedensityparametrization} are given by
\be
\ab{\mathcal{D}}=\frac{H^0_{\mathcal{D}}+\frac13 H^n_{\mathcal{D}}}{\mathscr{N}}\,,\,\,\,
a_\mathcal{D} = \frac{H^n_{\mathcal{D}}}{H^0_{\mathcal{D}}+\frac13 H^n_{\mathcal{D}}}\,,\,\,\,
b_\mathcal{D} = \frac{H^o_{\mathcal{D}}}{\mathscr{N}}\, .
\label{eq:Dabintermsofmatrixelements}
\ee
The angular dependence of each element in the density matrix is determined by Eq.~\ref{eq:chargedensityparametrization}.
For instance,  given the polarizations $\epsilon^\mu_{\pm} = \frac{1}{\sqrt{2}}(0,1,\pm i,0)$ and $\epsilon_0^\mu=(0,0,0,1)$,
the parity-even terms are obtained from $\vec{\epsilon}_\pm\cdot \vec{n}=  \frac{e^{\pm i\phi}}{\sqrt{2}}\sin\theta$ and $\vec{\epsilon}_0\cdot \vec{n}=\cos\theta$ while the nonvanishing contractions for the parity-odd term are given by $-i\epsilon^{ijk}\epsilon_{\pm,i}^*n_j\epsilon_{\pm,k} = \pm \cos\theta$, and $-i\epsilon^{ijk}\epsilon_{0,i}^*n_j\epsilon_{\pm,k} = \mp \frac{e^{\pm i\phi}}{\sqrt{2}}\sin\theta$.

Consider the left handed current injected via a positively charged $W$ boson at rest decaying to $u\bar{d}$, with the $\bar{d}$ having positive helicity due to the left coupling.
The total charge $\ab{\mathcal{D}}$, for the case of a conserved charge, is obviously determined to all orders by the charge of the operator.
The total charges are given by 
$\ab{\mathcal{E}} = m_W$, $\ab{\mathcal{Q}}=1$,  $\ab{\mathcal{I}}=1$ and $\ab{\mathcal{B}}=0$ for the energy, electric charge, isospin, and baryon number.

The $a_\mathcal{D}$ and $b_\mathcal{D}$ coefficients depend on the dynamics and are computed perturbatively.
At leading order these are simply determined by the different decay amplitudes $\mathcal{A}_h$. In the vector rest frame, these are simply proportional to the Wigner-d matrices,
$\mathcal{A}_h = g_V m_V e^{ih\phi}d^1_{1,h}(\theta_\star)$, with $d^1_{1,\pm 1}=\frac{1\pm\cos\theta_\star}{2}$ and $d^1_{1,0}=-\frac{1}{\sqrt{2}}\sin\theta_\star$,
where $\theta_*$ is the angle between the spin of the vector and the helicity-plus fermion in the vector rest frame. 

In the case of the energy correlator, one has $b_\mathcal{E}=0$ and $a_\mathcal{E}=-3/2$ \cite{PhysRevD.17.2298,PhysRevLett.41.1585,PhysRevD.19.2018}, which saturates the lower bound of \cite{Hofman:2008ar}.
In the case of the isospin, since both $\bar{d}$ and $u$ quarks carry the same quantum number, at leading order one has $b_\mathcal{I}=0$ and $a_\mathcal{I}=a_\mathcal{E}$.
The baryon number is more interesting, since while $\ab{\mathcal{B}}=0$, the individual quarks do carry baryon number and therefore a nontrivial distribution is predicted. One has $a_\mathcal{B}=0$ and 
$b_\mathcal{B}=-3B_{q}$ where $B_{q}=1/3$ is the absolute value of the baryon number of the individual quarks.
For the electric charge, the parity-even coefficients depend on the sum of charges of the light quarks which coincides with the sum of isospins, namely $a_\mathcal{Q}=a_\mathcal{I}$. The parity-odd coefficients depend on the difference of quark charges, and therefore $b_\mathcal{Q}=b_\mathcal{B}/2$.

At NLO, whenever the detector measures the charge of one of the quarks, two types of diagrams contribute, corresponding to the square of the real emission diagram and the virtual correction to the vertex. In dimensional regularization, both contain soft and collinear divergences that cancel in the sum. For the charge correlators, the cancellation is similar to the case of the inclusive cross section, since the detector acts homogeneously on the diagrams, simply fixing the quark direction. 
In the energy correlators, the cancellation of the collinear divergence only takes place after adding the third contribution coming from placing the detector on the gauge boson, as it is required due to energy conservation of the collinear splitting. This third contribution does not have virtual correction and it does not contain any soft divergence due to the extra energy insertion in the gauge boson's energy integral.
The vectorlike nature of the gluon coupling implies that the parity-odd structure in Eq.~\ref{eq:Hmunudecompositionfull} is not modified at one loop. The only contribution to $b_\mathcal{Q}$ and $b_\mathcal{B}$ is via the correction of the total rate.
The summary of the results is reported in Table~\ref{tab:nloresults}.
\begin{table}[]
        \renewcommand{\arraystretch}{1.40} 
        \addtolength{\tabcolsep}{7.00pt}
	\begin{tabular}{cccc}
		& $\ab{\mathcal{D}}$	& $a_\mathcal{D}$ & $b_\mathcal{D}$ \\ 
		$\mathcal{E}$ &	$m_W$ 	& $-\frac{3}{2}+ \frac{9\alpha_S}{2\pi}$ & 0 \\
		$\mathcal{Q}$ &	$1$ 	& $-\frac{3}{2}+  \frac{6\alpha_S}{2\pi}$ & $-\frac{3}{2}B_q\left( 1-\frac{\alpha_S}{\pi}\right)$ \\
		$\mathcal{I}$ &	$1$ 	& $-\frac{3}{2}+  \frac{6\alpha_S}{2\pi}$ & 0 \\
		$\mathcal{B}$ &	$0$ 	& 0 & $-3B_q\left( 1-\frac{\alpha_S}{\pi}\right)$ 
	\end{tabular}
	\caption{Coefficients of Eq.~\ref{eq:chargedensityparametrization} at one loop for the densities of energy $\mathcal{E}$, electric charge $\mathcal{Q}$, isospin $\mathcal{I}$ and baryon number $\mathcal{B}$ for a positively charged $W$-boson.}
	\label{tab:nloresults}
\end{table}

\subsection{Perturbative and non-perturbative corrections}

\begin{figure*}
	\centering
	\includegraphics[width=1\textwidth]{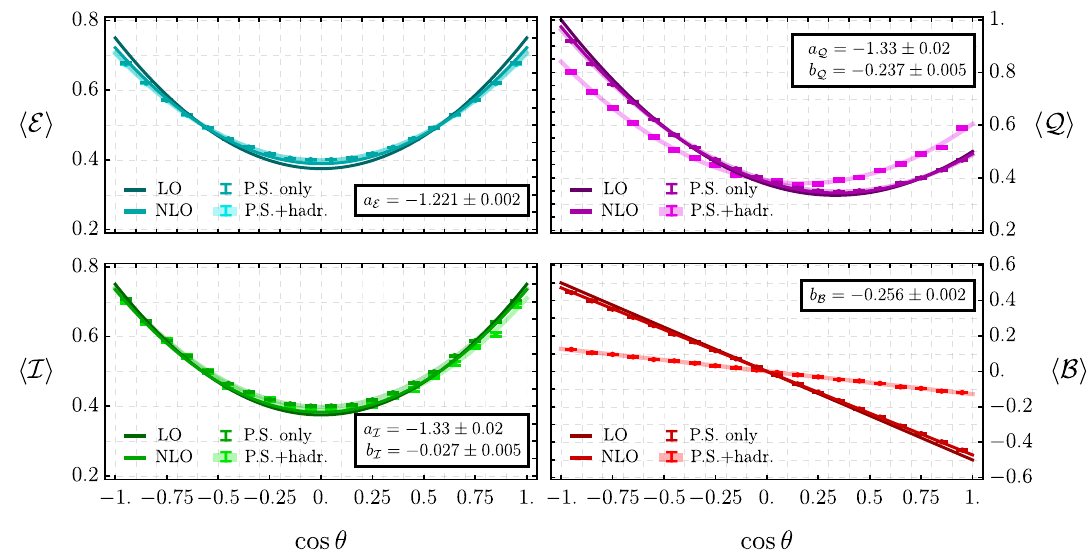}
	\caption{Density of QCD global charges for a $W$ boson decay, as a function of the angle $\cos\theta$ between the $W$ spin and the detector. From left to right and top to bottom, energy, electric charge, isospin and baryon number. The dark- and light-colored lines denote the LO and NLO prediction. The dark- and light- colored data points the \PT\, output with only parton shower (\textit{P.S. only}) and including hadronization (\textit{P.S.+hadr.}). In the inset, the fitted values for $a_\mathcal{D}$ and $b_\mathcal{D}$ of the \textit{P.S.+hadr.} sample.}
	\label{fig:Wppdecaychargedensities}
\end{figure*}

In the following we present a comparison between the analytical calculation and the numerical simulation. We simulate an $e^+ \nu_e$ collision in \MG\, \cite{Alwall:2014hca} and implement the parton shower and the hadronization via \PT\, \cite{Bierlich:2022pfr}. Given the left handed nature of the lepton current, the $W$ spin is aligned with the beam and only the plus helicity is produced, i.e. the production density matrix is $\rho^{(\ell\nu)}\propto \text{diag}(1,0,0)$, which implies that the $W$ spin axis is aligned with the beam axis and the only relevant decay is $\rho^W_{++}$, with $\theta$ being the angle between the positron's momentum and the detector.
The Cabibbo angle is set ot zero at this level since we are interested in the $u-\bar{d}$ chiral current, so no $s$-quark is injected by the operator.
The result is shown in Fig.~\ref{fig:Wppdecaychargedensities}. 
From left to right and top to bottom, the densities of energy, electric charge, isospin and baryon number as functions of the angle $\cos\theta$ between the $W$ spin axis and the detector. The dark- and light-colored lines denote the LO and NLO predictions, respectively. The dark- and light-colored data points the \PT\, output with only parton shower without hadronization (\textit{P.S. only}) and including hadronization (\textit{P.S.+hadr.}). We simulate a total of 2.5 million events, and the quoted uncertainties refer only to the statistical uncertainty.

In all cases the NLO calculation does provide an improvement with respect to the LO calculation and an accurate prediction of the density after the parton shower, as it is expected from IR-safe quantities, confirming the arguments of the previous section. 
For the energy correlator, the contribution from the gluon tends to flatten out the distribution, and the fitted $a_\cale$ parameter without the hadronization is $a_\cale=-1.232\pm 0.003$. 
The parity-even part of both isospin and electric charge distribution have the same coefficient at NLO, and compatible with the fitted values $a_\mathcal{I}=-1.335\pm 0.005$ and $a_\mathcal{Q}=-1.334\pm 0.005$.
Up to NLO, the isospin distribution is symmetric since $u$ and $\bar{d}$ carry positive isospin. At two loops a nonzero $b_\mathcal{I}$ receives a potential contribution from a nonplanar diagram. The parity-odd coefficient electric charge $b_\mathcal{Q}$ does receive a tree-level contribution, and the NLO correction pushes it to higher values compatible with $b_\mathcal{Q}=-0.475\pm 0.001$.
The baryon density is completely asymmetric and positive (negative) in the region with predominantly a quark (antiquark). 
The baryon asymmetry $b_\mathcal{B}$ is twice the electric charge one, and observed to be $b_\mathcal{B}=-0.939\pm 0.001$ with parton shower only.

The hadronization and weak interactions induce a series of corrections to the densities. First of all, isospin is not preserved by weak interactions, which induce $K^0-\bar{K}^0$ mixing. This implies an asymmetry between the $u$-quark and the $\bar{d}$-quark since up-quarks hadronize into charged kaons of definite isospins and reach the detector whereas down-quarks hadronize 
into neutral kaons, either $K_S$ decaying into pions or $K_L$ reaching the detectors. 
These effects generate an asymmetry, predicted to be $b_\mathcal{I}=-0.013\pm 0.002$ by the \PT\, simulation. Moreover, this also generates a total isospin charge larger that the $W$-boson's isospin, $\ab{\mathcal{I}}=1.023\pm 0.0007$.

The baryon asymmetry gets a large correction due to non-perturbative effects.  At the perturbative level, it is predicted that the baryon number carried by a quark and its collinear radiation is $\sim1/3$. This would suggest that the average baryon number observed tends to $1/3$. However, string formation spoils factorization between the collinear sectors,
and the asymptotic baryon number observed strongly depends on the details of how the QCD string breaks. The hadronization model used in \PT\, is based on the Lund string model \cite{Andersson:1983jt,SJOSTRAND1984469}, which controls the baryon production through the ratio between the rate of diquark-antidiquark production relative to the quark-antiquark production in the string breaking
\cite{ANDERSSON198245,PhysRevD.20.179,Andersson:1984af}. In Pythia, this is controlled by the parameter \verb|StringFlav:probQQtoQ|, and set to the default value of $0.081$ in Fig.~\ref{fig:Wppdecaychargedensities}, which leads to $b_\mathcal{B}=-0.128\pm 0.001$. 
In the range between $\sim0.04$ and $\sim0.16$ for the Pythia parameter, the asymptotic baryon asymmetry behaves close to linear, $b_\mathcal{B}\simeq -0.128\times \verb|StringFlav:probQQtoQ|/0.081$.

The electric charge receives non-perturbative corrections already mentioned.
Since pions have equal charge and isospin, measuring only pions would lead to a symmetric electric charge distribution. The asymmetry in the charge at the hadron level should be driven by the baryons and kaons. Indeed, the $\cos\theta<0$ region is populated by protons and neutrons with vanishing average isospin and positive average electric charge while the $\cos\theta>0$ region is populated by antiprotons and antineutrons. This only accounts partially for the electric charge asymmetry though since only baryons would lead to $b_\mathcal{Q}=b_\mathcal{B}/2$.
The rest comes from positively charged kaons predominantly generated at $\cos\theta<0$. 

\begin{figure}
	\centering
	\includegraphics[width=.48\textwidth]{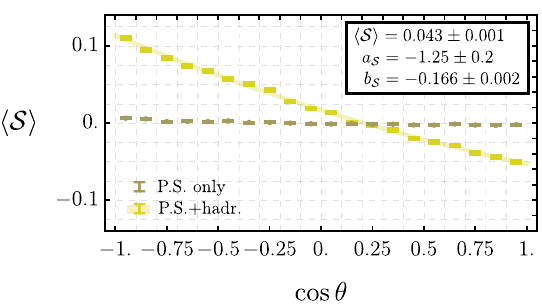}
	\caption{Density of strangeness generated nonperturbatively.}
	\label{fig:density_strang}
\end{figure}

Lastly, a nonzero strangeness density is generated nonperturbatively. 
The meachanism is exactly the one described for the isospin, as it is the production of kaons that are responsible for both effects. The generated strangeness charge of the current is predicted to be $\ab{\mathcal{S}}=0.043\pm 0.001$ by \PT\,, indeed twice the observed excess of isospin. The asymmetry is observed to be $b_\mathcal{S}=-1.66\pm 0.002$. 

This explains the aforementioned charge asymmetry since the different quantum numbers satisfy the relation $\ab{\mathcal{Q}_n}=\ab{\mathcal{I}_n}+\frac12\ab{\mathcal{B}_n}+\frac12\ab{\mathcal{S}_n}$ at a fixed direction $\vec{n}$, as long as no other quantum numbers are present.
Therefore, the different parity-odd parameters are related by $b_\mathcal{Q}\simeq b_\mathcal{I}+b_\mathcal{S}/2+b_\mathcal{B}/2$.
The density of strangeness is illustrated in Fig.~\ref{fig:density_strang}.

\section{Bounds on one-point correlators}

The coefficients in Eq.~\ref{eq:chargedensityparametrization} 
are subject to constraints stemming from the unitarity and the positivity of the energy. 
The spatial components of the hadronic tensor $H^{\mu\nu}$ in Eq.~\ref{eq:JDJcorrelator} form a positive definite matrix, as can be seen by inserting a complete set of states,
\bea
\int d^4x e^{ip\cdot x}\langle J^{i\dagger}(x) \mathcal{E}_n J^j(0)\rangle =\hspace{3.cm}\\ \nonumber
\sum_{\alpha,i\in \alpha}\delta^{(4)}(p_\alpha-p)
 E_i\delta^{(2)}(\Omega_i-\Omega_{n})\langle 0|J^{i\dagger} |\alpha\rangle\langle\alpha |J^j|0\rangle\,.
\eea
Therefore, the positivity of the energy implies that
the matrix spanned by the components of $\vec{J}$ is positive definite. 
Using the decomposition of Eq.~\ref{eq:Hmunudecompositionfull}, this positive definiteness implies two inequalities for the scalar terms independent of $\vec{n}$, and given by
\be
H_{\mathcal{E}}^0\,>\,0\,,\quad
H_{\mathcal{E}}^0 + H_{\mathcal{E}}^n\,>\,0\,.
\ee
This implies, obviously, that $\langle \mathcal{E}\rangle\,>\,0$, but also that $a_{\mathcal{E}}$ in Eq.~\ref{eq:chargedensityparametrization} and Eq.~\ref{eq:Dabintermsofmatrixelements} is bounded by
\be
-\frac{3}{2}\,\leq\, a_{\mathcal{E}} \,\leq\, 3\,,
\ee
with the lower and upper bound saturated when $H_{\mathcal{E}}^n=-H_{\mathcal{E}}^0$ and when $H_{\mathcal{E}}^0/H_{\mathcal{E}}^n = 0$, respectively.
This is the bound discussed in \cite{Hofman:2008ar}, obtained here as a simple consequence of Lorentz invariance, unitarity, and positivity of the energy.

The upper bound is saturated by two particle states with total angular momentum $j=1$ and projection $m= 0$, so the distribution is proportional to $1-\cos^2\theta$, while the lower bound is saturated by two particle states with $j=1$ and $m=\pm 1$, so the distribution is proportional to $1+\cos^2\theta$.
It is often stated \cite{Hofman:2008ar} that such cases are obtained by considering weakly coupled scalars and fermions, respectively, so that $J^\mu = \phi^*\overleftrightarrow{D}_\mu \phi$ and $J^\mu = \bar{\psi} \gamma^\mu \psi$.
However, the situation can be reversed and the upper bound can be saturated with fermions and the lower with scalars, if those are nonminimally coupled with $\mathcal{L}\supset \bar{\psi}\sigma^{\mu\nu} \psi F_{\mu\nu}$ and $\mathcal{L}\supset D^\mu \phi^* D^\nu \phi F_{\mu\nu}$.

From the discussion above it is clear that the same string of reasoning applies to any observable which is positive definite, in particular to a detector measuring the energy to any real power $k\geq 0$, $\mathcal{E}^k$, see \cite{Kravchuk:2018htv} for an equivalent statement for CFT. Consequently, one has that $H_{\mathcal{E}^k}^0\,>\,0$ and $H_{\mathcal{E}^k}^0 + H_{\mathcal{E}^k}^n\,>\,0$, which implies not only $\langle \mathcal{E}^k\rangle\,>\,0$, but also that $-\frac{3}{2}\,\leq\, a_{\mathcal{E}^k} \,\leq\, 3$ in exact analogy with the $k=1$ case.
Again, the bounds are saturated by two particle states with $m=\pm 1$ and $m=0$.

Since QCD has a mass gap, the restriction of $k\geq 0$ can be lifted and consider $k$ to be any real number as the measurement will be finite. However, QED corrections lead to soft photons and the observable formally diverges for negative $k$, even though one could assume a finite resolution. Different values of $k$ explore the theory in different regimes.
\begin{figure}
	\centering
	\includegraphics[width=.47\textwidth]{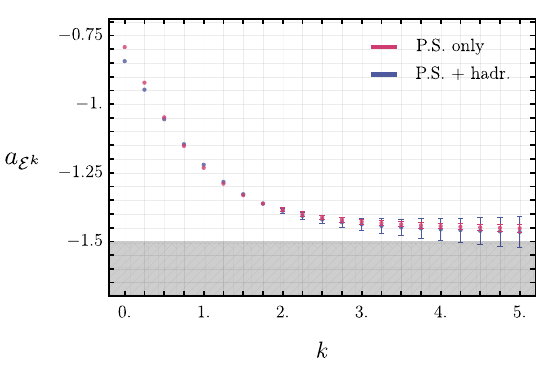}
	\caption{Coefficient $a_{\cale^k}$ as defined in Eq.~\ref{eq:chargedensityparametrization} as a function of $k$ for a vector current decaying into quarks. Red and blue denote \PT\, output without and with hadronization, respectively.}
	\label{fig:akplot}
\end{figure}
We provide in Fig.~\ref{fig:akplot} the coefficients $a_{\cale^k}$ from $k=0$ to $k=5$ extracted from a sample of $W$-boson decays as specified in the previous section. For $k=1$, $a_\cale$ indeed corresponds to the value in Table~\ref{tab:nloresults} and the distribution in Fig.~\ref{fig:Wppdecaychargedensities}. 
For large $k$, data tends to saturate the unitarity bound. This is due to the fact that in this region
collinear and hard splittings are suppressed and only 
events with few highly energetic particles plus a soft sector dominate, looking more and more like the tree-level process, corresponding to free fermions.
For smaller $k$, softer particles start to contribute and for $k=0$ the $\cale^k$ detector becomes the particle number detector and the distribution is the number density distribution.

\subsection{Relation with moments}

Beyond the positivity of $H_{\mathcal{E}^k}^0$ and of $H_{\mathcal{E}^k}^0 + H_{\mathcal{E}^k}^n$, which we denote either as $H(k)$ from now on for simplicity, there are a further infinite set of constraints that such quantities satisfy. 
They stem from the fact that both can be written as
\be
H(k)\,=\, \int d\Phi \rho(\Phi) (\mathcal{E}(\Phi))^k\,,
\ee
i.e. as integrals over some kinematic configuration $\Phi$, a positive distribution $\rho(\Phi)\geq 0$ and a positive function of such kinematics $\mathcal{E}(\Phi)\geq 0$, normalized in such a way that $\mathcal{E}(\Phi)\leq 1$, raised to some power $k>0$. This implies that the derivatives obey
\be
(-1)^N\frac{d^N H(k)}{dk^N}\,>\,0\, ,
\label{eq:completelymonotonicfunction}
\ee
and therefore $H(k)$ (i.e., both $H_{\mathcal{E}^k}$ and $H_{\mathcal{E}^k}^0 + H_{\mathcal{E}^k}^n$) is a completely monotonic function in $k$. 
The fact that $H(k)$ is a completely monotonic function means that its functional form is constrained. 
The problem we will be interested in is whether the measurement of $H(k)$ in some interval for $k$ can be used to extrapolate the function outside such interval.
We do this not by assuming a certain functional form for $H(k)$, but by finding the complete set of constraints that the values $\{H(k_1),\dots\}$, obtained by evaluating the function on a discrete set $\{k_1,\dots\}$, have to satisfy.

For $N=0$, Eq.~\ref{eq:completelymonotonicfunction} reduces to the positivity of $H(k)$. 
The constraints from $N>0$ can be efficiently posed by considering a discrete sequence $k,\, k+\delta,\, k+2\delta,\,\dots$. It can be shown 
\cite{ba7d9c4e-23b8-3532-8d1d-bfc038dcf811} (see also Chapter~IV of \cite{widder1941laplace}) that the set of constants 
$H(k),\, H(k+\delta),\, H(k+2\delta),\,\dots$ form a completely monotonic sequence,
\be
(-1)^N \Delta^N H(k+m\delta)\geq 0\, ,
\label{eq:completelymonotonicHk}
\ee
for any $N,m\geq 0$, and with $\Delta$ being the discrete derivative, $\Delta^0 H(k)=H(k)$ and $\Delta^n H(k) = H^{n-1}(k+\delta)-H^{n-1}(k)$. 
The full set of linear constraints imply that the set of $H(k+m\delta)$ can be identified with moments of a positive distribution, and therefore making an equivalence with the Hausdorff moment problem \cite{Hausdorff19211,Hausdorff19212}, see \cite{Bellazzini:2020cot} for a recent discussion in the context of scattering amplitudes. 
In particular, $H(k)$ is completely monotonic in $k$ if for any $\delta>0$ the following matrices $H^{(p)}$ with entries $(H^p)_{ij}=H(k+(i+j+p-2)\delta)$ are positive definite:
\be
\begin{split}
&H^0\succ0\,,\quad
H^1\succ0\, ,
\\[5pt]
&H^0-H^1\succ0\,,\quad
H^1-H^2\succ0.
\label{eq:Hankelmatricesconstraints}
\end{split}
\ee
The advantage of this formulation of the constraints on $H(k)$ is that a finite set of the nonlinear constraints of Eq.~\ref{eq:Hankelmatricesconstraints} is equivalent to a class of infinite set of linear constraints of Eq.~\ref{eq:completelymonotonicHk}.
Therefore the moment structure of $H(k)$ constrains the behavior of $\ab{\cale^k}$ and $a_{\cale^k}$ as a function of $k$ and Eq.~\ref{eq:Hankelmatricesconstraints} provides the complete, optimal set of constraints.

Given a set of moments, the space of higher moments allowed by the requirements in Eq.~\ref{eq:Hankelmatricesconstraints} is exponentially small.
We illustrate this point with the following. We extract $\ab{\cale^k}$ and $a_{\cale^k}$ for $k=1,1.25,\dots,3$ from a numerical simulation used in Fig.~\ref{fig:akplot}. The best fit points are used to extract $H^0_{\mathcal{E}^k}$ and $H_{\mathcal{E}^k}^0 + H_{\mathcal{E}^k}^n$ from the relations in Eq.~\ref{eq:Dabintermsofmatrixelements}. 
Note that $\mathscr{N}$ is an overall normalization independent of $k$ and does not affect the constraints.
Now we can set up a simple Semidefinite programming (SDP) routine that extremizes the allowed values for $k=3.25$ up to $k=5$, by requiring that the quantities $H^0_{\mathcal{E}^{k=1}},\dots,H^0_{\mathcal{E}^{k=5}}$ obey the constraints in Eq.~\ref{eq:Hankelmatricesconstraints}, and similarly for $H_{\mathcal{E}^k}^0 + H_{\mathcal{E}^k}^n$. We find that $\ab{\cale^{3.25}}$ is constrained at the level of $10^{-5}$ while $\ab{\cale^{5}}$ is determined within $2\cdot 10^{-2}$. Therefore, for $k$ close to the region where the parameters are known, the $\ab{\cale^{k}}$ are very efficiently constrained by the moment structure while larger $k$ have slightly more freedom. For instance, the intermediate $\ab{\cale^{4}}$ is constrained at the $10^{-3}$ level.
A similar observation applies to $a_{\cale^k}$, but with stronger relative constrains by an order of magnitude. In this case, for $k=3.25$ the bounds are at the $2\cdot 10^{-6}$ level, $10^{-4}$ for $k=4$ and $2\cdot 10^{-3}$ for $k=5$.

Ideally, the fit on $a_{\cale^k}$ and $\ab{\cale^k}$ should be performed together with the requirement of the moment constraints. While low values of $k$ are limited by systematic uncertainties, large values of $k$ are of easier experimental access but suffer of larger statistical uncertainties. 

 \begin{figure}
	\centering
	\includegraphics[width=1\linewidth]{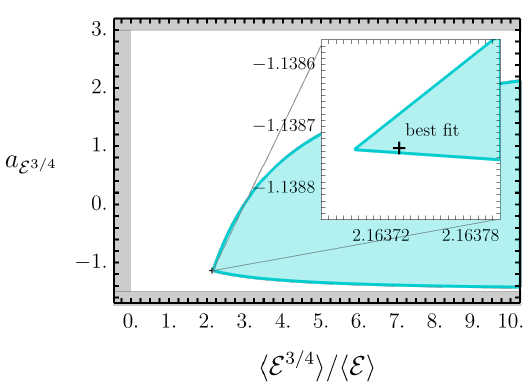}
	\caption{Theoretical constraints on $a_{\cale^k}$ and $\ab{\cale^k}$ for $k=0.75$ and fixing the values for $k\geq 1$ to their best fit values. 
	In grey the bounds $-\frac32\leq a_{\cale^k}\leq 3$ and $\ab{\cale^k}>0$. In teal, the allowed space from the moment structure. The actual best fit point sits strikingly close to the \textit{kink}.}
	\label{fig:plotSDP}
\end{figure}

While the constraints on higher moments are quite stringent, the constraints on lower moments are significantly looser. We consider a similar setup, where the moments for $k=1,1.25,\dots, 3$ are fixed to the best fit points, and we run an SDP that extremizes the allowed values for $k=0.75$. Running the SDP with only $H^0_{\mathcal{E}^{k=0.75}}$ as free parameter finds a lower bound, but the problem is unbounded by above. The same is true for $H^0_{\mathcal{E}^{k=0.75}}+H^n_{\mathcal{E}^{k=0.75}}$.
This is illustrated in Fig.~\ref{fig:plotSDP}, where we show in teal the allowed space in $a_{\cale^k}$ and $\ab{\cale^k}/\ab{\cale}$ for $k=0.75$ once the values for $k\geq 1$ are fixed to their best fit values. The lower bounds for $H^0_{\mathcal{E}^{k=0.75}}$ and $H^0_{\mathcal{E}^{k=0.75}}+H^n_{\mathcal{E}^{k=0.75}}$ leads to a sharp boundary for the allowed space. 
The fact that those have no upper bounds lead to the unbounded region in the right of the plot.
What is most striking, is that the actual extracted values for $k=0.75$ are extraordinarily close to the \textit{kink} of the allowed space.
This suggests that both $H^0_{\mathcal{E}^{k}}$ and $H^0_{\mathcal{E}^{k}}+H^n_{\mathcal{E}^{k}}$ have a power law behavior that is close to saturate the moment structure.

The phenomenological value of this class of constraints is that allows to use experimentally clean, IR-safe data for $k\gtrsim 1$, in order to constrain experimentally challenging, non-IR-safe data at low $k$. In particular, number density corresponds to $k=0$, which can be constrained by IR-safe observables with $k>0$.

We end with a speculative note: while from the arguments presented here one cannot rule out points in the bulk of Fig.~\ref{fig:plotSDP}, in particular points with arbitrarly large $\ab{\cale^{3/4}}$, in order to be in that region a dramatic change in $H(k)$ must occur. We find reasonable that there should exist extra constraints, possibly linked with full unitarity of the S-matrix, that would rule out such extreme behaviors, since this is what forbids large hierarchies among moments associated with scattering amplitudes \cite{Riembau:2022yse}. For instance, in theories with a mass gap $m_\pi$, $\ab{\cale^{k=0}}$ has the upper bound $Q/m_\pi$, with $Q$ being the hard scale.

\section{Conclusions and outlook}

In this \textit{Letter} we have explored the use of one-point correlators in order to understand the density matrix of conserved and non-conserved charges in QCD.
We observe that the one-point correlators of conserved charges are IR-safe perturbatively, which includes not only the energy density, but also the densities of electric charge, isospin and other flavors, and baryon number.
We compare the analytical result with the numerical simulation in Fig.~\ref{fig:Wppdecaychargedensities}. 
We remark that despite the simplicity of the observables, the densities carry an incredible amount of information on the underlying dynamics.
The functional form of the different densities is fixed by Lorentz invariance. 
We present a complete and optimal set of constraints that the densities of $\cale^k$ should satisfy. This allows to constrain the allowed space of $a_{\cale^k}$ and $\ab{\cale^k}$ as in Fig.~\ref{fig:plotSDP}. The fitted value from simulation is close to the kink, which indicates that the actual distribution is close to a power law.

The most interesting avenue is an experimental measurement of such densities, and the extraction of the parameters $\ab{\mathcal{D}}$, $a_\mathcal{D}$ and $b_\mathcal{D}$.
A promising context to perform such measurements at the LHC is in semi-leptonic $t\bar{t}$, as it offers a relatively clean sample of hadronically decaying $W$ bosons \cite{Kats:2015cna,Kats:2023gul} and the statistical sample of semileptonic $t\bar{t}$ is reaching the level of hadronic $Z$-decay at LEP. Combined with charm jet identification \cite{CMS:2021scf}, charm densities might be a target.

In the future, FCC-ee is expected to provide around $10^{13}$ $Z$-bosons and $10^8$ $W$-boson pairs \cite{FCC:2018byv}, which would allow to perform measurements of the different densities to an unprecedented precision.

In the same way the energy density can be used to extract the $W$ spin density matrix \cite{Ricci:2022htc},
the family of observables proposed here may boost the discriminating power and allow to better characterize a jet.

While focused on one-point correlators, the arguments presented apply to a subset of correlators beyond one-point. This opens up the space of calculable correlators, with a network of connections to other correlators via theoretical constraints and the operator product expansion. The theoretical and phenomenological exploration of this network may lead to new perspectives on collider physics.

\subsection*{Acknowledgments}

We thank Tim Cohen, Pier Monni, Lorenzo Ricci, Alba Soto and Alexander Zhiboedov for useful discussion and comments on our manuscript.
MS was supported by National Research Foundation of Korea under Grant Number NRF-2021R1A2C1095430.

\bibliography{bibs} 

\end{document}